%% file: CsMag_template.tex
\documentclass{IEEEcsmag}

\usepackage[colorlinks,urlcolor=blue,linkcolor=blue,citecolor=blue]{hyperref}
\expandafter\def\expandafter\UrlBreaks\expandafter{\UrlBreaks\do\/\do\*\do\-\do\~\do\'\do\"\do\-}
\usepackage{upmath,color}
\usepackage[super]{natbib}

\jvol{XX}
\jnum{XX}
\paper{8}
\jmonth{Month}
\jname{Publication Name}
\jtitle{Publication Title}
\pubyear{2021}

\begin{document}

\sptitle{Feature Article: Visualization Analytics Framework}

\title{idMotif: An Interactive Motif Identification in Protein Sequences}

\author{Ji Hwan Park$^*$ }
\affil{University of Oklahoma, Norman, OK, 73019, USA}

\author{Vikash Prasad}
\affil{University of Oklahoma, Norman, OK, 73019, USA}

\author{Sydney Newsom}
\affil{University of Oklahoma, Norman, OK, 73019, USA}

\author{Fares Najar}
\affil{Oklahoma State University, Stillwater, OK 74078, USA}

\author{Rakhi Rajan}
\affil{University of Oklahoma, Norman, OK, 73019, USA}

\markboth{THEME/FEATURE/DEPARTMENT}{THEME/FEATURE/DEPARTMENT}

\begin{abstract}\looseness-1This article presents a visual analytics framework, idMotif, to support domain experts in identifying motifs in protein sequences. A motif is a short sequence of amino acids usually associated with distinct functions of a protein, and identifying similar motifs in protein sequences helps to predict certain types of disease or infection. idMotif can be used to explore, analyze, and visualize such motifs in protein sequences. We introduce a deep-learning-based method for grouping protein sequences and allow users to discover motif candidates of protein groups based on local explanations of the decision of a deep-learning model. idMotif provides several interactive linked views for between and within protein cluster/group and sequence analysis. Through a case study and experts’ feedback, we demonstrate how the framework helps domain experts analyze protein sequences and motif identification.
\end{abstract}


\input{sec-introduction}
\input{sec-relatedwork}
\input{sec-task}

\input{sec-processing}

\input{sec-method}

\input{sec-casestudies}

\input{sec-discussion}

\input{sec-conclusion}

\section{ACKNOWLEDGMENTS}
This work was supported by the University of Oklahoma Data Institute for Societal Challenges Seed Funding Program.



\def\refname{REFERENCES}

\bibliographystyle{ieeetr}
\bibliography{template}

\begin{IEEEbiography}{Ji Hwan Park}{\,}is an assistant professor in the School of Computer Science at the University of Oklahoma, Norman, OK, 73019, USA. His research focuses on visualization, visual analytics, machine learning, and data science. He received his Ph.D. degree in computer science from Stony Brook University. He is a member of IEEE. He is the corresponding author of this article. Contact him at jpark@ou.edu.\vspace*{8pt}
\end{IEEEbiography}

\begin{IEEEbiography}{Vikash Prasad}{\,}is a Ph.D. student at the University of Oklahoma, Norman, OK, 73019, USA. His research focuses on the intersection of natural language processing, bioinformatics, visualization, visual analytics, and machine learning. He received his Master's degree from the Indian Institute of Technology-Indian School of Mines, Dhanbad. Contact him at vikash.k.prasad-1@ou.edu
\vspace*{8pt}
\end{IEEEbiography}

\begin{IEEEbiography}{Sydney Newsom}{\,} was a research assistant and a course instructor at the University of Oklahoma in Norman, OK, 73019, USA, during the time this work was performed. Her research interests included protein DNA interactions and genome biology. Sydney received a Ph.D. degree in biochemistry from the University of Oklahoma. Contact her at newsomsn@yahoo.com.
\vspace*{8pt}
\end{IEEEbiography}

\begin{IEEEbiography}{Fares Najar}{\,}is a Bioinformatics Research Scientist at the High-Performance Computing Center (HPCC) at Oklahoma State University, Stillwater, OK, 74078, USA.  His research interest is investigating different biological systems and visualization. He received his Ph.D. from the University of Oklahoma. Contact him at fnajar@okstate.edu.

\vspace*{8pt}
\end{IEEEbiography}

\begin{IEEEbiography}{Rakhi Rajan}{\,}is an Associate Professor in the Department of Chemistry and Biochemistry at the University of Oklahoma, Norman, OK, 73019, USA.  Her research focuses on characterizing the molecular mechanisms of CRISPR-Cas systems, the RNA-guided protein immune systems found in bacteria and archaea. She received her Ph.D. from the Ohio State University, Columbus, OH. Contact her at r-rajan@ou.edu.

\vspace*{8pt}
\end{IEEEbiography}
\end{document}

%% file: sec-introduction.tex
\maketitle
\chapteri{T}he sequences of proteins are integral to dictating their function. The specific order of amino acids, placement of a conserved set of amino acids at specific regions of a protein, and segregation of protein structure/sequence into subdomains within a multi-domain protein all signify functional importance. The protein sequences are conserved during the process of evolution such that amino acid properties across the length of the protein are preserved by amino acids with identical properties. Detecting conserved amino acid patterns within the whole protein sequence, termed motifs, and the position of each amino acid within the motif can signify the critical role of that motif towards a biological function. For example, the identification of motifs that are unique to a pathogenic bacterial protein can be a new strategy to treat bacterial infections.


Biochemists constantly depend on protein sequence analysis and visualization tools to analyze/identify important protein motifs and to design experiments to test the role of a motif.
While several methods have been proposed for automatically detecting motifs, these approaches can miss some motifs/locations or be computationally expensive~\cite{Pan:2019}. Thus, it is necessary to develop an interactive tool that allows domain experts to examine protein sequences and motifs for verification even though motifs are discovered automatically.      
Existing work for analyzing and visualizing biological sequences contributes to many areas, including identifying sequence differences~\cite{Nusrat:2019} and comprehending protein interaction~\cite{Furmanova:2020}. In addition, several methods focus on sequence consensus or an overview of sequences~\cite{Nusrat:2019}.
However, these tools don't allow users to interactively explore the characteristics of proteins that are segregated into different groups by comparing individual protein sequences or sequences that are representative of that particular group. Such interactive analysis will aid biochemists in discovering motifs that maybe present in protein sequences.      

To address this issue, we present idMotif to enable a user to get insights into protein sequences and identify motif candidates by observing sequence patterns. It is designed through an iterative discussion with a team of biochemists.  
Instead of automatically identifying motif candidates, we compute the possibility of each location as a motif candidate by using a deep-learning-based approach. This possibility is referred to as the saliency value and is visualized in idMotif. In our deep-learning model for motif identification, we utilized a pre-trained model to generate embeddings of protein sequences. We then fine-tune the model to predict groups of protein sequences. Lastly, we compute motif candidates by applying an interpretation method to the fine-tuned model. idMotif offers multiple linked views to present an overview and details of individual and groups of protein sequences as well as the saliency values of the sequences. It helps domain experts explore protein sequences, comprehend the characteristics of the sequences in different groups of protein sequences, and identify motif candidates. We conducted a case study and had expert feedback to demonstrate the effectiveness and usability of idMotif in identifying motif candidates. The main contributions of idMotif are summarized as follows:
\begin{itemize}
    \item We present an interactive visual analytics framework, idMotif, to assist domain experts in analyzing individual and groups of protein sequences and identifying motif candidates.
    \item We offer a deep-learning based analytical approach that fine-tunes a deep-learning model and applies an interpretation model to the fine-tuned model to discover motif candidates.
    \item We demonstrate the effectiveness of our framework with a case study and through qualitative user feedback from domain experts.
\end{itemize}



%% file: sec-relatedwork.tex
\section{RELATED WORK}

\subsection{Motif Discovery}
Several methods have been proposed to identify motifs automatically. DeepBind~\cite{Alipanahi:2015} uses a convolutional neural network (CNN) to identify nucleotide motifs. The authors convert the frequency of the four nucleotide bases of DNA sequences at each location to four 2D channels and predict the existence of each base at each location. iDeepS~\cite{Pan:2018} predicts RNA-protein binding sites by combining two CNNs and a Bidirectional Long Short Term Memory network (Bi-LSTM). In the method, the CNNs extract the features of the input sequences, and the Bi-LSTM is exploited to capture the long-term dependency between the extracted features. Recently, Yamada and Hamada~\cite{Yamada:2022} used Bidirectional Encoder Representations from Transformers (BERT) for forecasting the interaction between RNA sequences and RNA-binding proteins and analyzed the attention of BERT for interpreting the prediction. 
All these methods require manual sequence exploration for domain experts' verification because they can miss some motifs. idMotif aims to incorporate several interactive views, a fine-tuned pre-trained Transformer model, and an interpretation model in motif discovery to aid domain experts in detecting motif candidates.    

\subsection{Sequence Visualization}
There are various frameworks for visualizing and analyzing sequences in many areas~\cite{Guo:2022}.
These frameworks aid users in extracting patterns, discovering the relationship between inputs and outcomes of sequence models, and forecasting sequences.
For example, LSTMVis~\cite{Strobelt:2018} is a visual analytics framework for helping users understand hidden states in a recurrent neural network. The framework visualizes text with hidden state patterns similar to a specific text defined by a user. idMotif not only displays similar protein sequences to a representative protein sequence but also allows users to compare groups of protein sequences.

Some of the existing sequence visualization methods focus on comparing multiple biological sequences.
Multiple sequence comparisons reveal high similarity in sub-sequences/regions in two or more sequences, in terms of location, order, proximity, and orientation~\cite{Nusrat:2019}.  
For instance, Furmanová et al.~\cite{Furmanova:2020} developed an interactive visualization framework for protein complex exploration, comparison, and filtering at various levels of detail.
invis~\cite{Demiralp:2013} is a visual analysis tool to explore projected RNA sequence data and compare sequences.    
Strobelt et al.~\cite{Strobelt:2016} proposed the Vials for understanding alternative mRNA sequences by comparing single and groups of samples. Additionally, the tool can be used for quality control of data.
Unlike these methods, the focus of our work is to comprehend the characteristics of individual and groups of protein sequences and identify motif candidates. 

In addition, motif visualization can be used for the prediction of protein structures and functions.
Existing motif visualization methods are helpful in building consensus on sequences or comparing nucleotides or amino acids at each location to show relationship patterns between locations within multiple sequences~\cite{Nusrat:2019}.
While these motif visualization methods focus on identifying motifs in a single type/group of protein sequences, idMotif enables users to build sequence groups and identify and compare motifs of each protein group based on their features extracted from a deep learning model.
   

%% file: sec-task.tex
\section{BACKGROUND AND TASKS}
This section presents how protein sequences were obtained and the target tasks of idMotif. We held several meetings with domain experts in bioinformatics and biochemistry to define the target tasks.  

\subsection{Protein Sequence Acquisition}
Most protein sequence data is obtained through nucleic acid sequencing and processing as follows:
1.	A nucleic acid, a fundamental building block, is extracted from a cell or tissue sample. The nucleic acid sample is the input for a sequencing platform, which utilizes a sequencing-by-synthesis approach to get multiple short reads covering the input nucleic acid sample. The output is many overlapping short nucleotide sequence reads.\\
2.	Sequence assembly software is then used to assemble the input of short overlapping sequencing reads into contigs (longer and less fragmented sequences) using either de novo techniques or a similar organism’s assembled nucleic acid sequence as a reference template. The output is a long and intact nucleic acid sequence, as opposed to ones that are short and fragmented. The assembled output nucleotide sequence may be as complete as a whole microbial genome or whole mammalian chromosome.\\
3.	The assembled sequences are deposited into a sequence database, and submitted sequences have undergone a combination of automated and manual processing. One aspect of the genome annotation pipeline is that it identifies open reading frames and lists the genes found in that nucleic acid sequence. The software also translates the gene sequences into protein sequences comprised of amino acid sequences.\\

\subsection{Tasks}
We adopted Sedlmair et al.'s nine-stage framework~\cite{Sedlmair:2012} for problem discovery and visualization design and implementation.
We collaborated closely with two domain experts who have worked in bioinformatics and biochemistry for over 20 years.
We had multiple discussions with them at different stages of development, each lasting 30 minutes to 1 hour.
During our discussions with the domain experts, they expressed a need for a tool to help identify motifs in groups of protein sequences. Specifically, they wanted a tool that could explore both individual and groups of protein sequences and discover motif candidates related to specific groups of protein sequences. Once identified, they could then validate the results through experiments in their lab.

To meet this need, we identified several tasks that the tool should be able to perform based on the discussions:

\paragraph{\textbf{T1 Classify groups of protein sequences:}} Some protein sequences can have unique characteristics that distinguish them from other types of protein sequences. The tool should enable domain experts to explore and discover groups of protein sequences with common characteristics.

\paragraph{\textbf{T2 Identify motif candidates:}} Each group of sequences can have short sequence patterns, motifs, associated with distinct functions within the group. Domain experts are interested in identifying motif candidates in each group.   

\paragraph{\textbf{T3 Compare different groups of protein sequences:}} 
There can be distinctive sequence patterns for different groups of protein sequences. Comparing these patterns can aid domain experts in understanding the basis for the separation of protein sequences as distinct groups.   

\paragraph{\textbf{T4 Comprehend the similarity within each group:}} In some cases, protein sequences in the same group can share only a few common amino acids yet still belong to the same type. It is crucial to identify the common patterns within each group. Additionally, protein sequences may be misclassified due to a clustering method. The similarity computation within a group helps domain experts discover misclassification and correct it. \\


%% file: sec-processing.tex
\section{MOTIF DISCOVERY}
\begin{figure}[t]
  \centering
  \includegraphics[width=\linewidth]{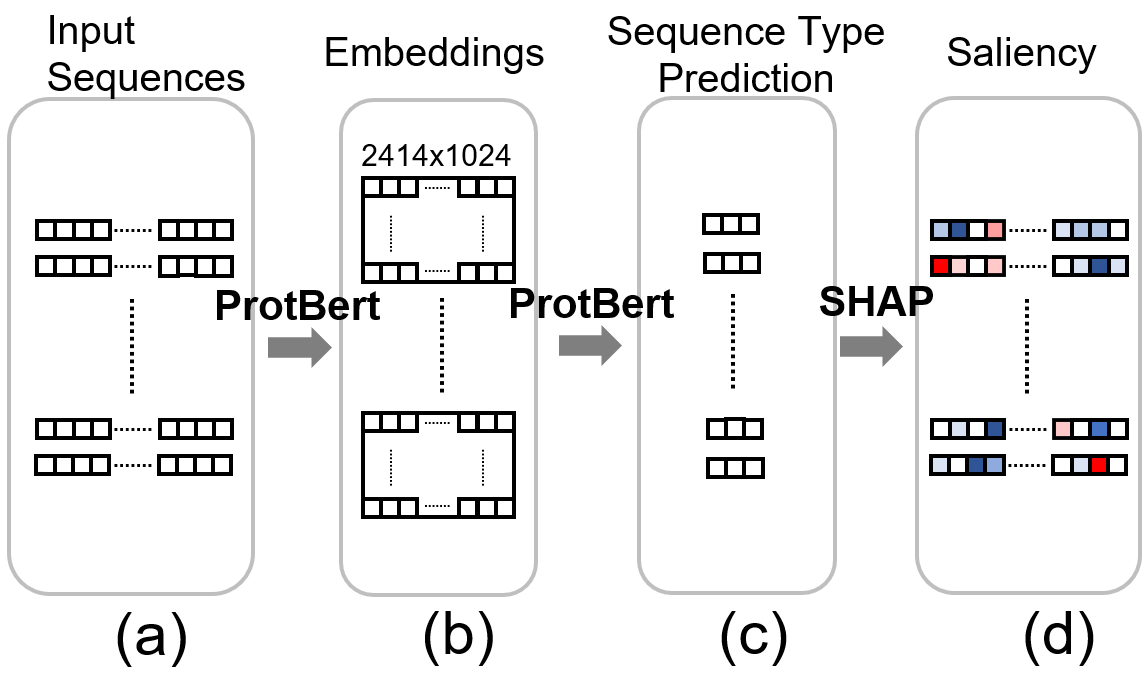}
  \caption{%
    A pipeline of the proposed motif identification method. (a) Given protein sequences, (b) we use a pre-trained model (ProtBert~\cite{Elnaggar:2020}) to generate embeddings for the sequences. Next, we (c) fine-tune a Transformer model (ProtBert) for sequence group prediction and then (d) apply a local explanation method (SHAP) to the sequences. %
  }
  \label{fig:processing}
\end{figure}

One of the goals of idMotif is to identify motif candidates in protein sequences. To achieve this goal, we developed a deep-learning-based motif identification method. The method consists of two steps: 1) training a deep-learning model to capture the characteristics of groups of protein sequences and 2) using a local explanation method for understanding the contribution of amino acids in the sequences of a particular group. Fig.~\ref{fig:processing} illustrates a pipeline of our identification method. 

\subsection{Sequence Prediction} \label{embedding}
To extract the characteristics of groups of protein sequences, we convert input protein sequences to embeddings and apply a deep-learning model to the embeddings.
A CNN and recurrent neural network (RNN), including long short-term memory (LSTM) and gated recurrent unit (GRU), can be used to classify protein sequences. While CNNs focus on local structure, Transformer-based models capture more global context than CNNs~\cite{Charoenkwan:2021}. 
Capturing global context is important in our task because motif candidates representing protein sequence groups can be long sequences comprising several hundreds of amino acids.  
In addition, although RNN-based models can capture long-term dependency in sequences similar to Transformer-based models, Transformer-based models have shown better performance than an RNN model and a CNN model in biological sequence prediction~\cite{Charoenkwan:2021}.  
Thus, in this work, we employed a Transformer-based model, ProtBert~\cite{Elnaggar:2020} that is a deep language model specifically designed for protein sequences, for protein sequence classification (Fig.~\ref{fig:processing}(b)).

ProtBert, a large language model, consists of both local and global representations, and has been trained on a large corpus of protein sequences. 
The amino acids have been treated as tokens, and ProtBert produces a vector representation with a length of 1024 for each amino acid token. The length of protein sequences can be varied, so first, we find the maximum length of protein sequences in our dataset, $Len_{max}$. To ensure consistent input size, we padded each protein sequence to the maximum length ($Len_{max}$), resulting in a standard size of $1 \times Len_{max} \times 1024$ for each protein sequence.   
 The use of ProtBert allowed us to obtain a powerful vector representation for each amino acid token in our protein sequence dataset, facilitating downstream analyses.\\
Subsequent to the generation of protein sequence embeddings using ProtBert, we conducted a fine-tuning process on the embedding dataset utilizing the ProtBert model (Fig.~\ref{fig:processing}(c)). The objective of this fine-tuning process was to create a model that can accurately differentiate and classify each protein sequence based on its underlying features.\\
During classification, the Transformer model outputs confidence probabilities for each group, indicating the model's belief in the presence of that particular class/group in the protein sequence. These probabilities reflect the model's prediction and serve as the basis for generating explainers.

\begin{figure}[t]
  \centering
  \includegraphics[width=\linewidth]{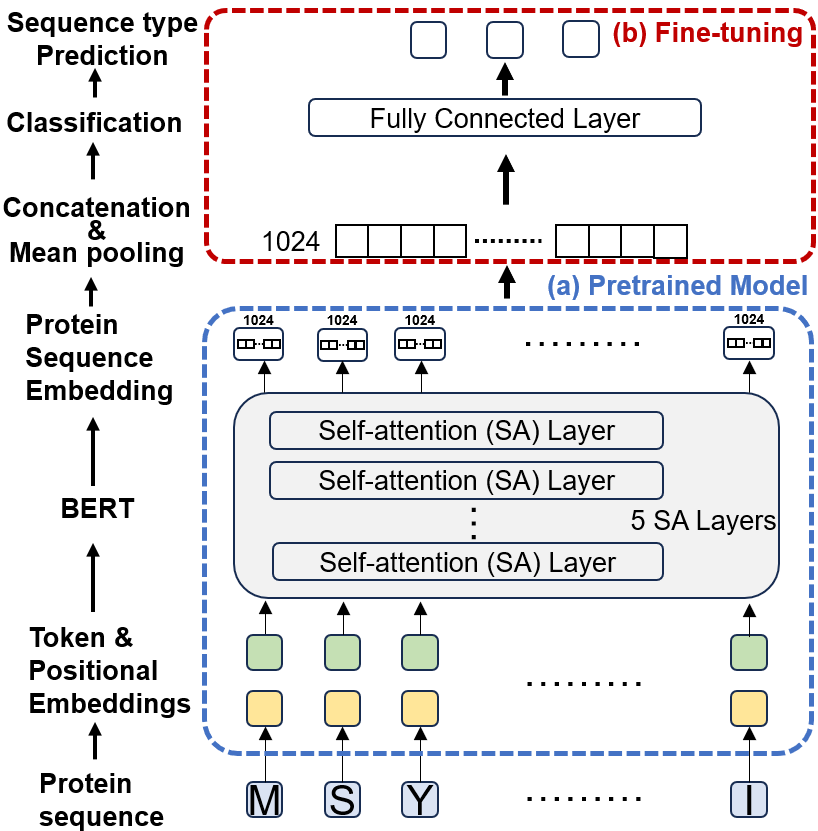}
  \caption{%
  	An architecture of our protein sequence prediction model, including (a) a pre-trained ProtBert and (b) a fine-tuning process.
  }
  \label{fig:protbert}
\end{figure}

\begin{figure*}[t]
  \centering
  \includegraphics[width=\linewidth]{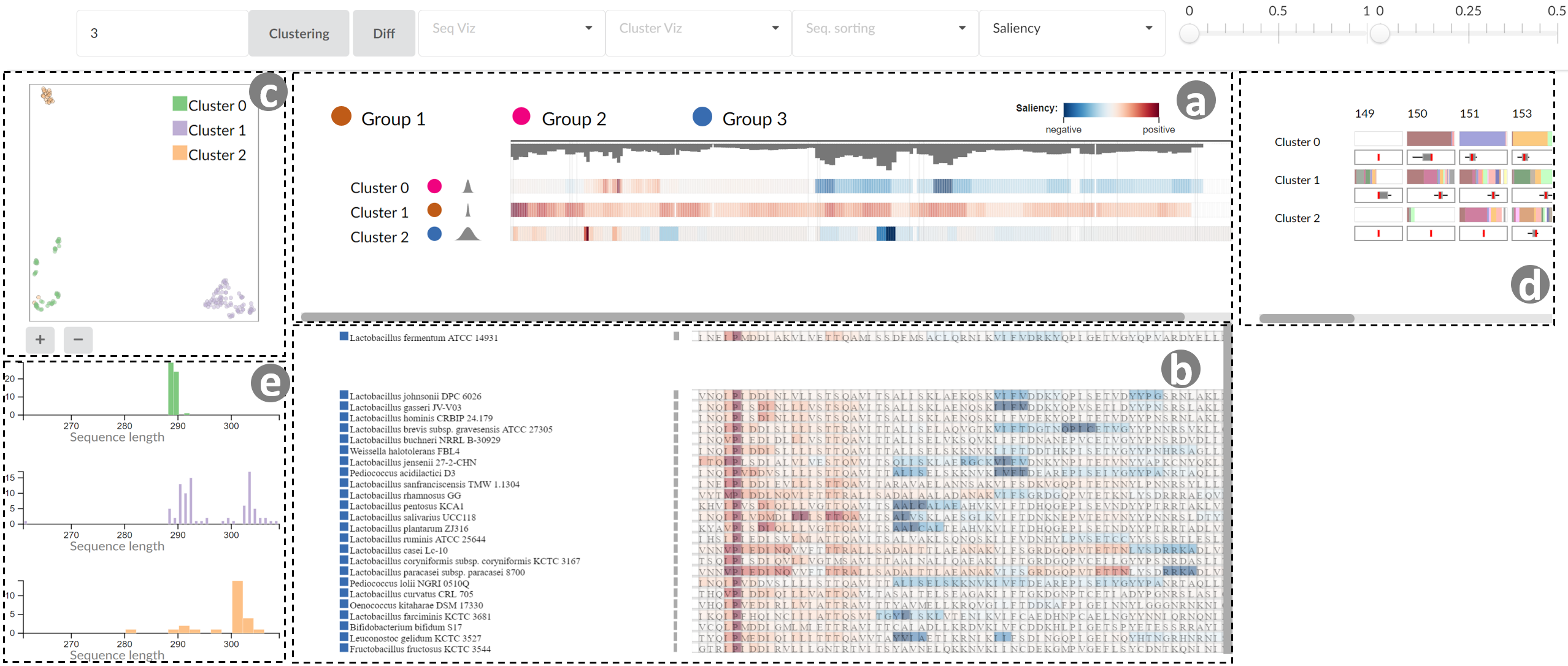}
  \caption{%
  	idMotif contains five linked views. (a) The Cluster view shows an overview of clustered protein sequences. (b) The Sequence view presents the details of each protein sequence. (c) The Projection view displays the similarity of protein sequences. (d) The Motif view displays the details of a selected region in the Cluster view for discovering motifs. Lastly, (e) the Distribution view visualizes the length distribution of protein sequences in a selected cluster. %
  }
  \label{fig:overview}
\end{figure*}

\paragraph{Model Architecture.} The ProtBert can generate features/embeddings for the input protein sequences (Fig.~\ref{fig:protbert}(a)). We begin by processing an incoming sequence of length $L$, which is composed of $L$ amino acids. The initial step involves tokenizing and positional encoding, which transform each amino acid into a vector. A pre-trained ProtBert model, which includes a stack of five self-attention layers with 16 attention heads, is applied to the vectors. This process generates protein sequence embeddings. 

We then fine-tune the ProtBert model to classify protein sequences (Fig.~\ref{fig:protbert}(b)). In fine-tuning, we first concatenate the embeddings and apply mean pooling. It generates a fixed-length (1024) vector. Subsequently, a fully-connected layer operates on the vector, followed by a softmax operation. 
The possible output sequence types of the fine-tuning part are the number of protein sequence groups (e.g., 3 for Cas1 dataset). In this fine-tuning process, we use a dataset with labels. The fine-tuned model can be used for similar protein sequence datasets. 
This process containing 105M parameters constitutes our comprehensive motif candidate identification pipeline.

\subsection{Motif Identification}
Existing work showed that deep learning models have a tendency to concentrate their prediction on traditional motifs~\cite{Lanchantin:2017}.
To understand which location our deep learning model focuses on, we deployed a popular local explanation model, SHAP~\cite{Attanasio:2023}, which can be applied to any deep learning model (Fig.~\ref{fig:processing}(d)).
Several local explanation techniques have been proposed to aid in comprehending machine learning models, including LIME, SHAP, and Integrated Gradients. SHAP stands out as it can accurately identify significant features in Transformer-based classifiers~\cite{Attanasio:2023}.
We apply SHAP to protein sequences. Internally, SHAP first performs tokenization on the input sequences, which converts the sequences into a sequence of tokens. It then generates a set of perturbations by creating modified versions of the input sequences, where each modification involves replacing a single token with a special "mask" token. SHAP uses a variant of the Shapley values algorithm to compute the contribution of each token to the model's prediction. The algorithm computes the difference in the model's output when a token is present compared to when it is replaced with the mask token. The contributions of all tokens are then aggregated to compute an overall explanation for the model's prediction. By analyzing these contributions, we can identify the specific parts of protein sequences, motifs, that contribute to their distinct grouping. In this work, we refer to these contributions as saliency values.

%% file: sec-method.tex
\section{IDMOTIF DESIGN}
To explore and analyze input protein sequences to perform the identified target tasks, we provide several linked views, as shown in Fig.~\ref{fig:overview}: the Projection view, the Cluster view, the Sequence view, the Motif view, and the Distribution view. In the Cluster view, the Sequence view, and the Motif view, we apply a global alignment~\cite{Edgar:2004} to protein sequences for comparing the sequences. Additionally, a user can highlight specific amino acids based on their saliency values by selecting a range of saliency values.

\paragraph{Workflow.}
In a typical workflow, users first examine the Projection view to identify groups of protein sequences. They then move to the Clustering view to compare each group, where certain regions are highlighted as potential motif candidates by a system. Additionally, users can understand the distribution of protein sequence lengths in the Distribution view, which may characterize the groups. At this stage, users may discover a motif candidate(s) based on the saliency values of representative sequences. In some cases, users explore protein sequences in each group in the Sequence view to correct misclassified sequences or to further refine a motif candidate. Finally, users confirm a motif candidate by selecting a region corresponding to a motif candidate sequence and inspecting the selected region's amino acids and saliency values in the Motif view.

\begin{figure}[t]
  \centering
  \includegraphics[width=\linewidth]{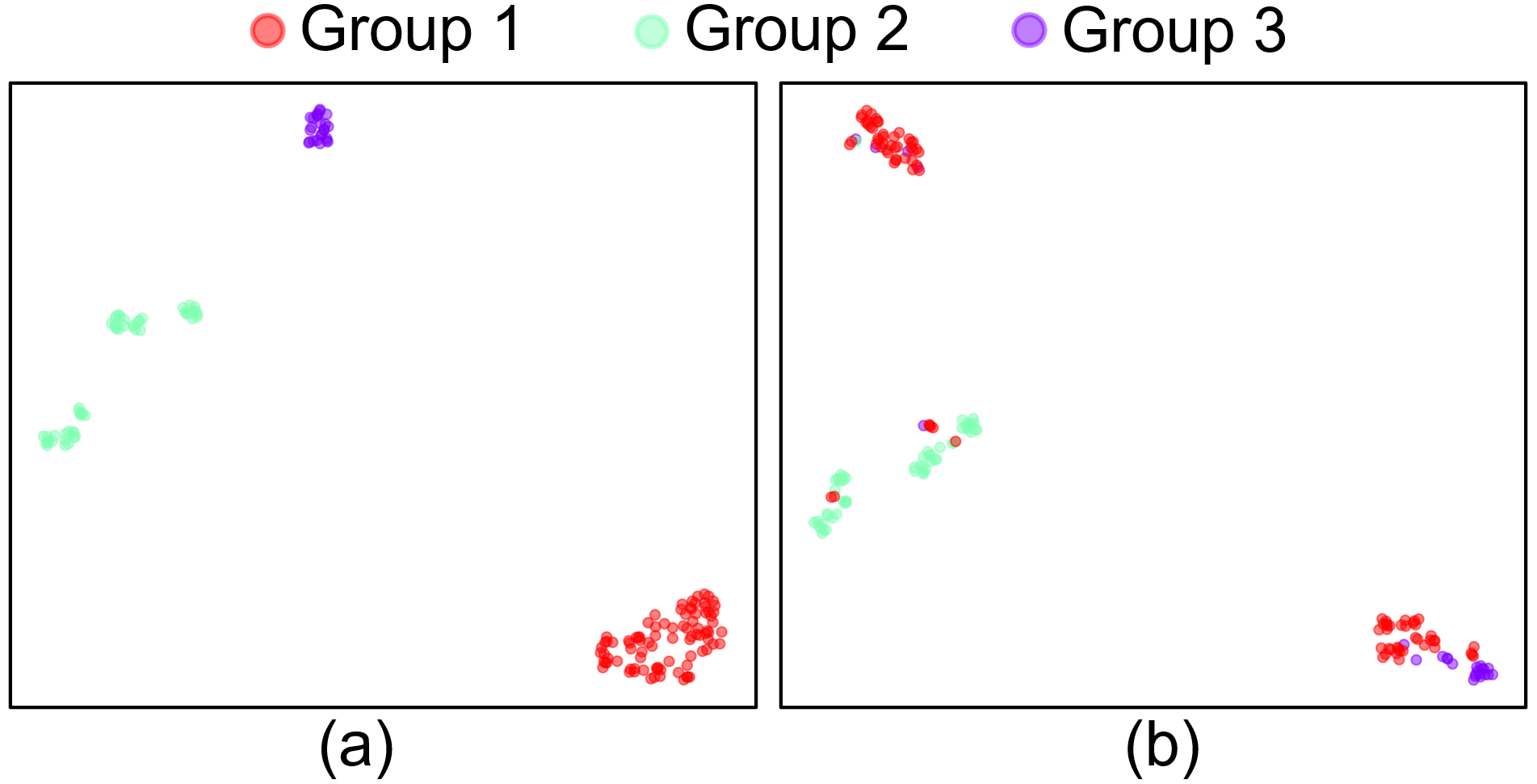}
   \vspace{-5 mm}
  \caption{%
  Comparison of two different projected data using UMAP: (a) the saliency values generated from SHAP analysis of the fine-tuned model, and (b) the embedding directly from the pre-trained ProtBert. Different colors indicate different groups of protein sequences. %
  }
 
  \label{fig:projection_input}
\end{figure}

\subsection{Projection View}
The Projection view provides the user with an overview of protein sequences based on their saliency values. In the Projection view, we project the saliency values of each protein sequence to a 2D scatter plot, where each circle represents a protein sequence. 
The Projection view uses the saliency values generated from SHAP analysis of the fine-tuned model as an input for the view instead of the embedding directly from the pre-trained ProtBert because it encodes more information and can better separate the protein sequences (Fig.~\ref{fig:projection_input}).  

This scatter plot assists the user in identifying and grouping similar sequences together easily (\textbf{T1}). There are several dimensionality reduction methods, and we applied UMAP~\cite{McInnes:2018} to the saliency values because it can preserve both the local and global structure of the input data.

After creating the scatter plot, we apply a clustering method to protein sequences (circles) in the plot to help the user find protein sequences with similar saliency values. By default, K-medoids clustering is used for partitioning the data points. 
The center point of each cluster will be displayed as a representative sequence for each cluster in the Cluster view. K-medoids clustering requires the number of clusters as an input parameter. The user interactively changes the number of clusters and finds an optimal value by exploring the results of clusters in the Projection view, the Cluster view, and the Sequence view.
If the user is not satisfied with the result of K-medoids, they can employ another clustering method.
Additionally, the users can select circles and create a new cluster manually by brushing. The selected circles are highlighted in the Sequence view, helping the user compare them.
We color each circle to its corresponding cluster in the Projection view. 
Fig.~\ref{fig:overview}(c) illustrates an example of the Projection view.

\begin{figure}[t]
  \centering
  \includegraphics[width=\linewidth]{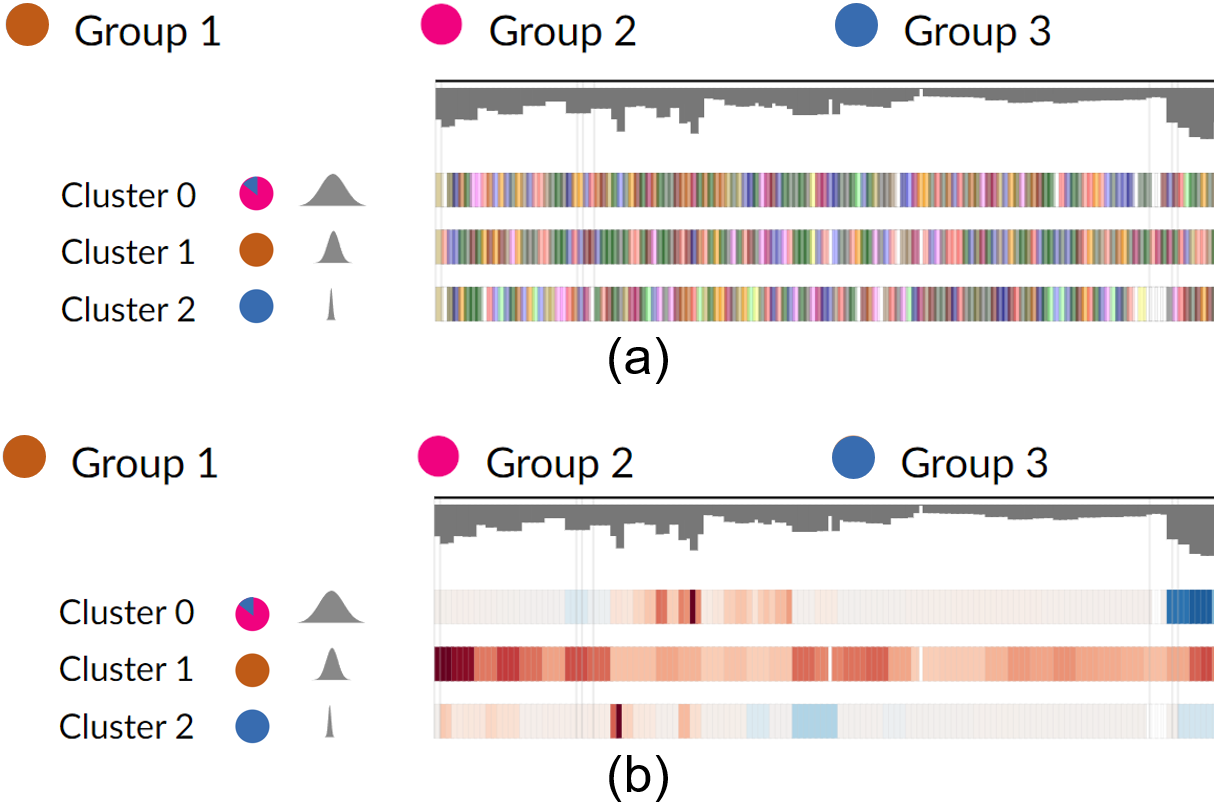}
  \vspace{-7 mm}
  \caption{%
    An example of the Cluster view. The Cluster view can visualize two types of information: (a) the type of amino acids and (b) the saliency values of amino acids. %
  }
  \label{fig:clusterview}
\end{figure}
 
\subsection{Cluster View}
The Cluster view allows the user to compare the similarities and differences between clusters based on their representative protein sequences (\textbf{T3}). Once the user finds the optimal number of clusters, a representative protein sequence for each cluster is visualized in the Cluster view, as illustrated in Fig.~\ref{fig:clusterview}. The Cluster view displays the sequences by using horizontal bars, where each row displays a representative protein sequence of each cluster, and each column/bar represents each sequence location. The color of each bar can encode the types of amino acids (Fig.~\ref{fig:clusterview}(a)) or the saliency values of amino acids (Fig.~\ref{fig:clusterview}(b)). After consulting with domain experts, we chose a color scheme for amino acid sequences that is widely used in existing protein sequence visualization tools.
It may be misinterpreted that there is a relationship between clusters and sequences, as both have identical color schemes. However, our primary objective was to utilize distinct colors in both clusters and sequences. 
Additionally, we use a color scheme for clusters in the views located on the left side of idMotif (the Projection view and the Distribution view), and a color scheme for sequences in the rest of the views on the right side of idMotif. This separation reduces the likelihood of any confusion or misinterpretation.

On the left side of the view, we display three types of information: 1) a cluster/group identified by our clustering method in the Projection view, 2) the ratio of the protein sequence groups assigned by domain experts (e.g., Group 1, Group 2, and Group 3 in Cas1 protein sequences~\cite{Orden:2017}), and 3) the variation of protein sequences in each cluster based on their saliency values. A pie chart shows the ratio of the sequence groups. The pie chart assists the user in understanding the uniformity of protein sequences in a cluster, i.e., whether there is a single protein group in the cluster. If there are multiple protein groups in a cluster, the user can further divide them by exploring sequences in the cluster in the Sequence view. Based on the discussion with domain experts, we display both clusters and protein groups. This is because domain experts wanted to see how accurate a clustering method is and what types of protein sequences the method misclassified. A Gaussian curve displays the variation of protein sequences in each cluster, where a standard deviation of the curve is the standard deviation of Euclidean distances between the center point and other data points of each cluster in the Projection view. Moreover, upon selecting a sequence, the user can assess the presence of either the same or different amino acids at each location.

At the top, bars with inverted $y$-axis represent the accumulated saliency values, where the \textit{x}-axis is the location of each amino acid, and \textit{y}-axis indicates the accumulated saliency values at each location. 
The accumulated saliency value at a location $i$ is computed as follows:
\begin{equation}
    Acc_i = \sum_{j=1}^{N}\|{S_{ij}}\|
\end{equation}
where $S_{ij}$ is a saliency value of $j$th protein sequence at $i$ location, and $N$ is the number of protein sequences. 
This chart enables the user to pinpoint significant regions within the clusters, which could potentially be considered as motif candidates.
In addition, the user is allowed to select a specific range of saliency values, and the corresponding locations will be highlighted.

We have observed that some columns in the Cluster view do not contain any amino acids or relevant information. For these instances, we removed the empty columns and indicated their deletion with lines. This streamlines the sequence comparison process and offers a concise overview of the representative sequences.


\paragraph{Design Rationale.}
Initially, we designed the Cluster view similar to the top part of the Sequence view, where we summarized the frequencies of the amino acids at each location for each cluster. However, we found that users couldn't easily understand the patterns among groups when motif candidates did not have the perfect consensus (e.g., one amino acid is prevalent in one location, but still many other amino acids exist at the same location within a group.). Identifying distinctive patterns or motif candidates for each cluster is one of our main tasks (T2), and the potential motif candidate locations can have similar amino acid and saliency value patterns. These observations motivated us to display a representative protein sequence for each cluster. We found that domain experts prefer horizontal bars to other shapes, including circles, because a group of bars with similar values/colors naturally form a region (rectangle), so they can easily discover them.  

\begin{figure}[t]
  \centering
  \includegraphics[width=\linewidth]{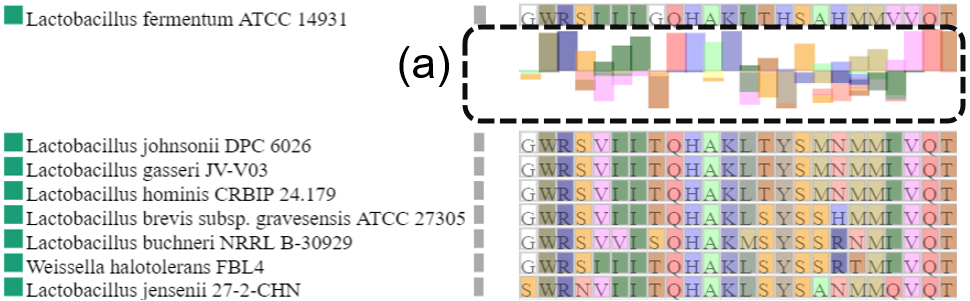}
  \caption{%
    An illustration of the Sequence view. (a) A stacked bar below a representative protein sequence is shown above the baseline to represent the frequency of amino acids matching the representative protein sequence at a specific position. For amino acids that do not match, their frequency is represented with bars stacked below the baseline.%
  }
  \label{fig:sequence_view}
\end{figure}

\subsection{Sequence View}
The user can choose a specific cluster in the Cluster view, and the details of all protein sequences in the chosen cluster are then visualized in the Sequence view (Fig.~\ref{fig:sequence_view}). This allows the user to comprehend the similarity and dissimilarity of protein sequences in the selected cluster (\textbf{T4}). The main part of the Sequence view displays the details of each sequence in a selected cluster, where each column indicates amino acid location, and each row represents a protein sequence.  The color of each cell corresponds to the types of amino acids and their saliency values, the same as the Cluster view.

The user can sort rows in the Sequence view based on embedding similarity, saliency similarity, and sequence similarity. The embedding similarity $ES^i_{r,j}$ is computed between a representative protein sequence $S^i_r$  and a protein sequence $S^i_j$ in a selected cluster $i$, as follows:
\begin{equation}
    ES^i_{r,j} = \frac{Emb(S^{i}_{r})\cdot Emb(S^i_j)}{\| Emb(S^{i}_{r}) \| \|  Emb(S^i_j)  \|} \quad {S^i_r\neq S^i_j} \textrm{ and } {S^i_j \in S^i} 
\end{equation}
where $Emb(x)$ is an embedding of a protein sequence $x$, and $S^i$ represents all the protein sequences in a cluster $i$.  To calculate saliency similarity, we substitute embedding with saliency values in Equation (2). For the sequence similarity, we count the number of the matching amino acids at each location between $S^i_r$ and $S^i_j$. 

idMotif also enables the user to manually select a different protein sequence in a cluster to serve as the representative protein sequence of the cluster. 
Similar to the Cluster view, the user can compare the representative protein sequence to other protein sequences in a cluster by highlighting the same or different sub-sequences.  
At the top, idMotif utilizes stacked bars to display the frequencies of the amino acids in a selected cluster (Fig.~\ref{fig:sequence_view}(a)), similar to $invis$~\cite{Demiralp:2013}. A stacked bar is shown above the baseline to represent the frequency of amino acids that match the representative sequence value at a specific location. Amino acids that don't match are represented with bars stacked below the baseline. On the left side, we display the group and name of each protein sequence. Additionally, the similarity between a representative protein sequence and each protein sequence is visualized by using a gray rectangle, where the width of the rectangle indicates the degree of similarity based on the chosen similarity computation method.

\subsection{Motif View}
The Motif view visualizes detailed information about specific locations selected in the Cluster view. This includes the frequencies of amino acids and the distribution of saliency values. The view shows the characteristics of selected sub-sequences and helps the user determine whether it is a motif candidate~(\textbf{T2}).
For each cluster, the view provides two types of information: horizontally stacked bars showing the amino acid frequencies (top) and a boxplot displaying the saliency value distribution across all input sequences (bottom) at the selected locations.
Figs.~\ref{fig:overview}(d) and~\ref{fig:motif_view} illustrate examples of the Motif view. 

 
\subsection{Distribution View}
\begin{figure}[t]
  \centering
  \includegraphics[width=0.9\linewidth]{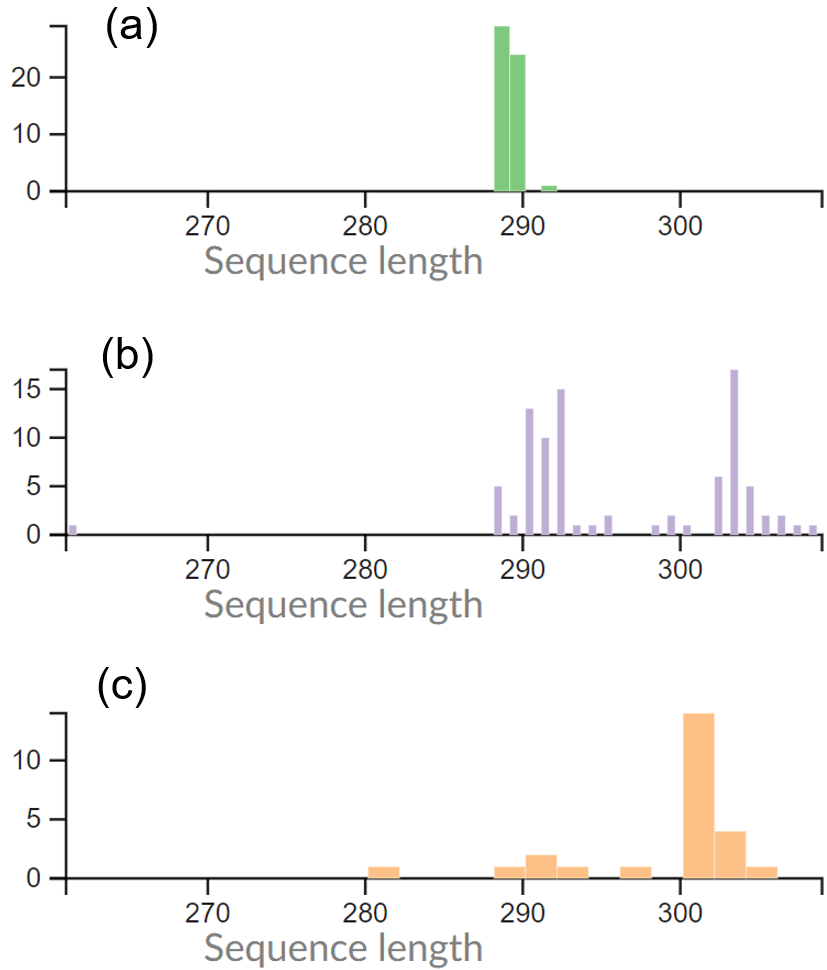}
  \vspace{-3 mm}
  \caption{%
   The distribution of sequence lengths for three groups of protein sequences: (a) Cluster 0, (b) Cluster 1, and (c) Cluster 2. %
  }
  \label{fig:infoview}
\end{figure}

A protein group can be associated with the length of its sequence, but it's not the only factor. To understand this relationship between groups of protein sequences and their length distribution, we display the length distribution of protein sequences for each cluster as histograms. In the Distribution view, users can compare length distributions across clusters easily. The Distribution view is updated when the user adds or removes a cluster in the Projection view.
Fig.~\ref{fig:infoview} shows the distribution of sequence lengths for three groups of protein sequences.

%% file: sec-casestudies.tex
\section{EVALUATION}
We evaluate the effectiveness of our motif identification method and present a case study as validation for idMotif. Furthermore, we conducted informal interviews with two domain experts in bioinformatics and biochemistry and collected their feedback.



\subsection{Dataset - Cas1 Sequences}
Two domain experts have utilized publicly available data to analyze protein sequence motifs present in proteins associated with type II-A clustered regularly interspaced short palindromic repeats (CRISPR)-CRISPR-associated (Cas) systems~\cite{Orden:2017}. 
CRISPR-Cas systems are adaptive immune systems that protect bacteria and archaea from invading genomes such as phages and plasmids. 
In this study, we focused on Cas1, a protein that belongs to the type II CRISPR-Cas systems, which plays a critical role in the adaptation stage of the CRISPR-Cas systems. Cas1, along with Cas2 protein, inserts pieces of foreign DNA into the CRISPR locus of bacterial or archaeal genome as ``spacer'' sequences. These spacer sequences are then utilized to generate CRISPR RNAs (crRNAs), which help the CRISPR-Cas systems to sequence-specifically recognize and eliminate foreign genetic materials in subsequent encounters. The domain experts analyzed 167 protein sequences and segregated each protein into three distinct groups~\cite{Orden:2017}.

 \begin{figure}[t]
  \centering
  \includegraphics[width=\linewidth]{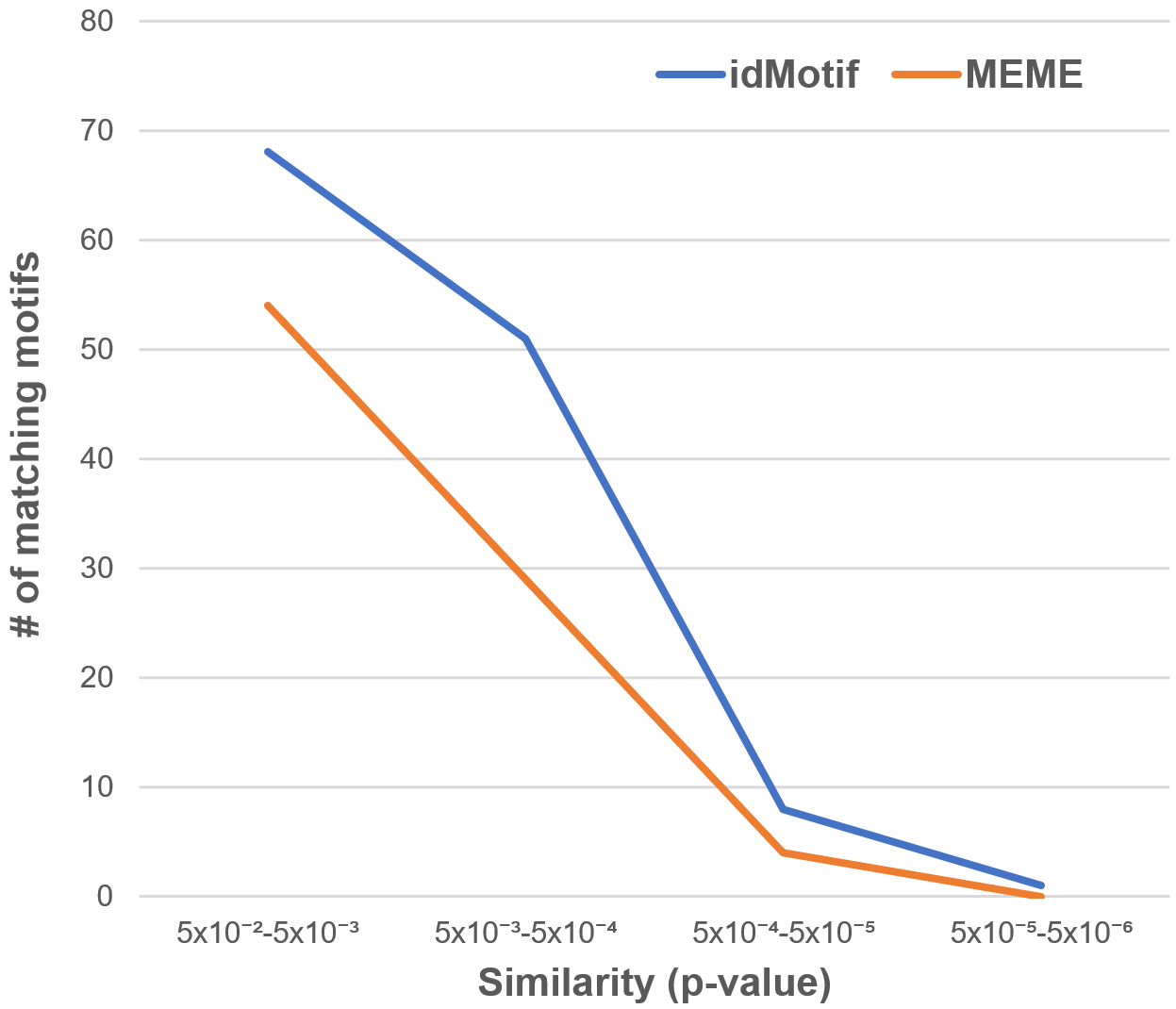}
  \vspace{-3 mm}
  \caption{%
  Accuracy comparison. We compared the accuracy of motif identification between our method and MEME by counting matching motifs with known motifs~\cite{Gupta:2008} within specific similarity ranges (p-values).
  }
  \label{fig:motif_accuracy}
\end{figure}

\begin{figure*}[t]
  \centering
  \includegraphics[width=0.98\linewidth]{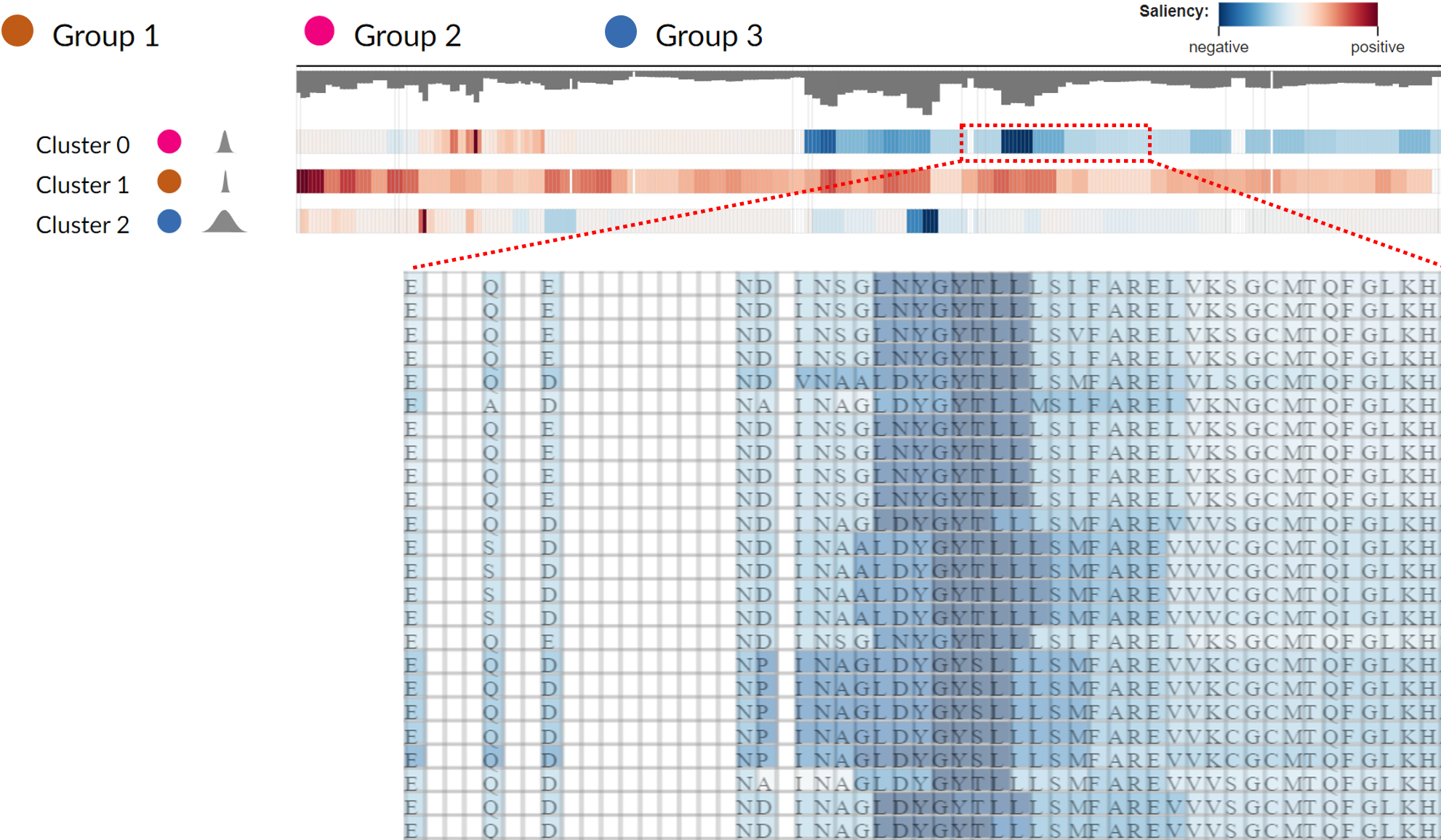}
  \vspace{-3 mm}
  \caption{%
  	An expert selected (a) a region in Cluster 0 in the Clustering view, and they verified that (b) protein sequences corresponding to the region have similar saliency values in the Sequence view. %
  }
  \label{fig:motif}
\end{figure*}

\subsection{Accuracy of Motif Identification}
We assessed the performance of motif identification using the Cas1 Dataset. Specifically, protein sequences belonging to each group (Group 1: 87 sequences, Group 2: 55 sequences, and Group 3: 25 sequences) were submitted to idMotif for identifying sequence features. To evaluate the accuracy of our motif identification method, we used a motif comparison algorithm~\cite{Gupta:2008}, which measures the similarity between the discovered motifs by our method and known motifs. The comparison algorithm generates a p-value, which indicates the probability of a random motif with the same width as the target having an optimal alignment with a match score as good as or better than the target. The smaller p-value means a detected motif is more accurate. Additionally, we computed p-values of motifs identified by a popular motif discovery algorithm, MEME~\cite{Machanick:2011}. In the experiment, we discarded motifs with large p-values (p-value > 0.05). We counted the number of matching motifs within specific ranges of p-values. Our result shows that idMotif detects motifs more accurately than MEME (Fig.~\ref{fig:motif_accuracy}).

\subsection{Case Study}

The experts began by exploring the Cas1 dataset to identify groups of protein sequences that shared similar characteristics in the Projection view (\textbf{T1}). They discovered three groups of protein sequences in the view (Fig.~\ref{fig:overview}(c)). The experts then inspected each group to identify any misclassified protein sequences (\textbf{T4}). In one group, Cluster 0, they found two protein sequences that were different from other sequences in the group based on the sequence similarity score. 
The group that these sequences were assigned by idMotif was different compared to the group assigned by domain experts, confirming that they were misclassified, and the experts corrected the group accordingly. 
Next, they analyzed the similarity among the groups by examining the amino acid patterns of the representative protein sequence of each group in the Cluster view. They found that there were distinctive patterns among the groups (Fig.~\ref{fig:clusterview}(a)). The experts also analyzed the distribution of protein sequence lengths of each group because each group might have a different distribution. They found that one group has relatively short sequences (Cluster 0, Fig.~\ref{fig:infoview}(a)), and another group has a distribution with relatively long sequences (Cluster 2, Fig.~\ref{fig:infoview}(c)), while Cluster 1 has a mix of both lengths of Clusters 0 and 2 (Fig.~\ref{fig:infoview}(b)). This helped the experts identify the unique pattern of each group (\textbf{T3}).

Additionally, they explored the saliency values of the representative sequences. They observed that a region in a group has a negative contribution to their classification. To confirm that this region has similar contributions across protein sequences within each group, they also analyzed the saliency values of the sequences in each cluster (\textbf{T4}). They compared a representative sequence to other sequences in the Sequence view and found that protein sequences corresponding to the region have similar saliency values (Fig.~\ref{fig:motif}).

\begin{figure}[t]
  \centering
  \includegraphics[width=\linewidth]{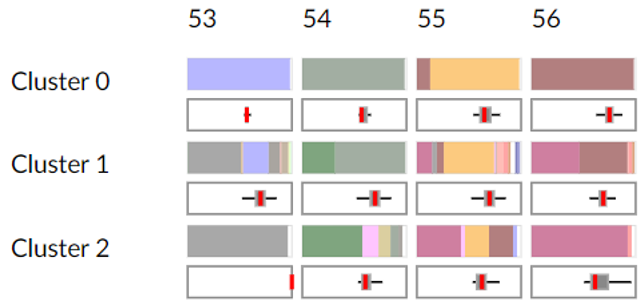}
  \vspace{-3 mm}
  \caption{%
   An example of motif identification in the Motif view. Some amino acids from location 53 to location 56, were prevalent in Cluster 0 and Cluster 2, respectively. %
  }
  \label{fig:motif_view}
\end{figure}

Lastly, they analyzed the saliency values of the protein sequences and pinpointed specific parts of the sequences that could potentially serve as motifs in the Cluster view. They achieved this by selecting regions with high accumulated saliency values (\textbf{T2}). Subsequently, they scrutinized each selected region in the Motif view to verify its potential. For instance, they found that some amino acids in a selected region, from location 53 to location 56, were prevalent in Cluster 0 and Cluster 2, as illustrated in Fig.~\ref{fig:motif_view}. Therefore, the experts confirmed that sequences corresponding to this region from Cluster 0 and Cluster 2 are motif candidates for Cluster 0 and Cluster 2.


\subsection{Expert Feedback}
We presented idMotif to two domain experts in bioinformatics and biochemistry and collected their feedback. They had experience with analyzing and visualizing biological data, including protein sequences. We held hour-long interviews.
First, we described the overall goals of the proposed tool and demonstrated the different interactive visual components of our platform to them. With a brief explanation, they were able to gain insights from each view. We then let the experts explore the system for 20 minutes freely. They used the Cas1 dataset. During the exploration, they analyzed the data and asked questions. After the exploration, the experts gave their feedback on our idMotif tool. 

 Overall, the feedback was positive, with the experts expressing satisfaction with the tool's ability to explore protein sequences and identify motif candidates. They particularly appreciated the Projection view, allowing them to quickly identify groups of protein sequences. One expert noted that the Sequence view was helpful in detecting outliers in the sequences. They also appreciated the Sequence view, as the deep-learning-based system effectively identified motifs. One expert mentioned that displaying saliency values in the Sequence view provides an additional classification level beyond existing alignment and classification methods.

In the Clustering view, the experts were able to identify and differentiate the unique characteristics and locations of each cluster. They found the tool helpful in comparing regions across different groups, providing insights into the differences among them. The Sequence view also aided in this understanding. Additionally, the Distribution view was deemed especially useful when analyzing proteins with varied sequence lengths. 
The experts expressed interest in using the tool to classify proteins based on whether they are from pathogenic or non-pathogenic bacteria and identify motifs that distinguish pathogenic vs. non-pathogenic bacterial proteins. 
Finally, they suggested adding displays for other parameters related to amino acids, such as hydrophobicity scales and transfer free energy, to enhance the tool's usefulness.

%% file: sec-discussion.tex
\section{Discussion}
\subsection{Impact on Motif Identification}

Motif identification for groups of protein sequences involves two steps: clustering and motif discovery. Most existing methods or tools using Matlab, R, or Python libraries or web applications (e.g., MEME~\cite{Machanick:2011}) that domain experts use in their daily practices for these tasks create text files or static plots. Moreover, these tools/methods require checking the result of each step manually and running a motif identification method for each detected group separately.
During the interview with domain experts, they expressed that as compared to these tools/methods, idMotif is a powerful tool with multiple linked views for clustering protein sequences, investigating different aspects of groups of protein sequences, and discovering motif candidates interactively. They also mentioned that idMotif provides an overview and the details of groups of protein sequences and motif candidates, which are useful for discovering outliers and refining motif candidates for groups of protein sequences.

\subsection{Limitations}
Although our proposed framework is effective for identifying motifs in protein sequences, it has some limitations. Firstly, it may not be suitable for other genetic data, such as DNA and RNA sequences, as it uses a pre-trained model specifically for protein sequences, ProtBert. To address this issue, we plan to explore other pre-trained models that are appropriate for different types of genetic data.

Secondly, the maximum length of protein sequences in the Cas1 dataset is 310, which can make visualizing and exploring large datasets with long sequences challenging, despite the various overview and filtering methods we provide. To solve this issue, we will investigate alternative approaches, such as hierarchical grouping/clustering, to offer a better overview of the data.

Lastly, in the fine-tuning process, to compute saliency values of amino acids, idMotif requires ground truth labels for protein sequences, i.e., all the protein sequences should be classified into one of the target protein groups. Unfortunately, obtaining ground truth labels is not always possible in some datasets. However, after the fine-tuning process is done on a protein dataset, the fine-tuned model can be applied to similar protein sequence datasets without labels.

%% file: sec-conclusion.tex
\section{CONCLUSION}
We introduced idMotif, a visual analytics framework for exploring protein sequences and discovering motif candidates interactively. Through the use of a deep-learning model and a local explanation model, we were able to accurately compute the contributions of amino acids and offer several linked views for analyzing individual and groups of protein sequences. 
Based on the presented case study and qualitative feedback from domain experts, we conclude that idMoitf is able to assist domain experts in understanding the characteristics of individual and groups of protein sequences and the relationship between protein sequences and motif candidates. DNA and RNA sequences consist of a chain of nucleotides, and each nucleotide can have one of the four bases, similar to there being 20 amino acids in protein sequences. Thus, we can generalize our framework to DNA and RNA sequences by replacing ProtBert with a pre-trained model for DNA/RNA sequences. In the future, we plan to use our framework for DNA/RNA sequences to discover their motifs.
